\documentclass[aps,prd,reprint,superscriptaddress,nofootinbib,longbibliography]{revtex4-2}
\usepackage{amsmath,amssymb,bm,mathtools,slashed}
\usepackage{booktabs}
\usepackage[colorlinks=true,linkcolor=black,citecolor=black,urlcolor=black]{hyperref}
\hypersetup{pdftitle={Weyl curvature and ultrasoft renormalization of the static QCD potential},pdfauthor={Michael Wilson}}
\newcommand{\as}{\alpha_s}
\newcommand{\CF}{C_F}
\newcommand{\CA}{C_A}
\newcommand{\TF}{T_F}
\newcommand{\Rz}{\mathcal R_z}
\newcommand{\WW}{\mathcal W}
\newcommand{\dd}{\mathrm d}
\newcommand{\tr}{\operatorname{tr}}
\newcommand{\eps}{\epsilon}
\newcommand{\MS}{\overline{\mathrm{MS}}}
\begin{document}
\title{Weyl curvature and ultrasoft renormalization of the static QCD potential}
\author{Michael Wilson}
\email{mkwilson3@ualr.edu}
\affiliation{Department of Physics, University of Arkansas at Little Rock, Little Rock, Arkansas, USA}
\begin{abstract}
We derive the leading gravitational correction to the ultrasoft renormalization of the static QCD potential. In a Ricci-flat background the electric Weyl tensor introduces a $t^{-2}$ term in the short-time chromoelectric correlator. The curvature-dependent divergence of the dipole self-energy is then linear rather than cubic in the singlet--octet energy gap, and occurs at order $\alpha_s^2$, two orders earlier than in the flat-space sector. For $W=C_{\mu\alpha\nu\beta}r^\mu u^\alpha r^\nu u^\beta$ and $V_s^{(W)}=c_W W/r$, we obtain
\[
\nu\,\partial_\nu c_W=\frac{C_FC_A}{3\pi}\alpha_s^2.
\]
We calculate the compensating soft infrared pole and the finite one-loop matching coefficient in an optical dimensional subtraction scheme. Independent physical-metric calculations reproduce the chromoelectric singularity and the finite light-flavor term. The leading regulated infrared overlap cancels in the matching, and the soft and ultrasoft reference constants transform oppositely under an evanescent curvature rescaling. In the perturbative adiabatic regime the static energy acquires an anisotropic $\alpha_s^2\ln\alpha_s$ term. The tidal field thus changes the perturbative order at which ultrasoft gluons contribute logarithmically to the static color interaction.
\end{abstract}
\maketitle

\section{Introduction}
The static quantum chromodynamics (QCD) potential is a Wilson coefficient rather than an observable by itself. Its infrared sensitivity is canceled by the ultraviolet sensitivity of dynamical gluons at the ultrasoft scale. This separation, made explicit in potential nonrelativistic QCD (pNRQCD) \cite{Brambilla2000,Brambilla2005,Pineda2012}, accounts for the first infrared divergence of the flat-space static potential at order $\as^4/r$ \cite{Appelquist1978,StaticIR,PinedaSoto2000}. The ultrasoft dipole self-energy carries three powers of the singlet--octet energy gap.

In a Ricci-flat background the static interaction acquires a dependence on the electric part of the Weyl tensor. The same curvature enters the propagation of the ultrasoft gluons. We consider the resulting correction to the chromoelectric dipole self-energy and its implications for the perturbative hierarchy of the static energy.

Gravitational corrections to electromagnetic two-body interactions have been calculated in weak-field geometries \cite{Caicedo2011}. Curved-space gluodynamics has also been studied using low-energy gluon--dilaton models \cite{Gaete2008}. At the level of perturbative amplitudes, gravitational couplings of quarks and gluons are encoded in the stress-tensor--current--current correlator \cite{Armillis2010,Coriano2025}; the corresponding massless quantum electrodynamics (QED) curvature effective action is known \cite{Donoghue2015}. The separation of soft and ultrasoft scales has also been studied in thermal pNRQCD \cite{Thermal2008}. Here the correction arises in the local singularity of the field-strength correlator rather than from thermal populations. We match a static source pair in an external Ricci-flat physical metric and determine the renormalization of its singlet potential by the chromoelectric dipole interaction. In particular, the soft Wilson-loop contributions and their ultrasoft subtraction are intrinsic to the static matching problem.

The static effective theory is formulated in Sec.~\ref{geom}. Section~\ref{tree} establishes the tree-level interaction, and Sec.~\ref{hard} computes the one-loop polarization and connected Wilson-line contributions. In Sec.~\ref{running} we derive the ultrasoft anomalous dimension, perform the infrared subtraction, and obtain the matched potential. Section~\ref{physferm} gives an independent physical-metric calculation of the light-flavor term. We discuss the resulting perturbative hierarchy in Sec.~\ref{discussion}.

\section{Static effective theory in a tidal field}
\label{geom}
We set $\hbar=c=1$, use signature $(-,+,+,+)$ and the curvature convention $R^\rho{}_{\sigma\mu\nu}=\partial_\mu\Gamma^\rho_{\nu\sigma}-\partial_\nu\Gamma^\rho_{\mu\sigma}+\cdots$. In a parallel tetrad along a central geodesic let $E_{ij}=R_{i0j0}$. For a Ricci-flat, purely electric Weyl field, $E^i{}_i=0$ and $R_{0ijk}=0$. At quadratic Fermi order, neglecting curvature derivatives, the Lorentzian metric takes the frozen form
\begin{align}
 ds^2&=-N^2dt^2+\gamma_{ij}\dd x^i\dd x^j,\\
 N&=1+\frac12E_{ij}x^ix^j,\qquad
 \gamma_{ij}=\delta_{ij}-\frac13 R_{ikjl}x^kx^l.
 \label{fermi}
\end{align}
Two static sources are placed at $\bm x_\pm=\pm\bm r/2$. The time coordinate $t$ equals proper time at the midpoint, not at either source. No thermal state or stationary Schwarzschild interior is assumed. We match at a fixed point of the central geodesic, treating the curvature as an external slowly varying field at the soft scale. The scale assumptions are
\begin{equation}
 m_Q\gg r^{-1}\gg \max(\sqrt{\|E\|},\Lambda_{\rm QCD}),
 \quad r\,\|\nabla E\|\ll\|E\|,
 \label{hierarchy}
\end{equation}
with weak coupling $\as(r^{-1})\ll1$. In a nonstationary background, the frozen soft coefficient also requires $r\Omega_E\ll1$, where $\Omega_E=\|D_t E\|/\|E\|$ measures the local rate of variation (away from zeros of the curvature; more generally the corresponding derivative expansion must be small). For a physical ultrasoft reference approximation, a distinct stronger hierarchy is specified in Sec.~\ref{running}. The mass $m_Q$ enters the static limit only as a scale; $n_f$ light flavors are massless. We set $\CF=(N_c^2-1)/(2N_c)$, $\CA=N_c$, $\TF=1/2$, and $g^2=4\pi\as$.

After Wick rotation of the frozen metric, the closed rectangular contour has temporal length $T$ and spatial connectors joining the sources at both ends. Denote its fundamental-representation Wilson loop by
\begin{equation}
 \WW(T,r)=\frac1{N_c}\tr\,\mathcal P\exp\left(ig\oint_{\mathcal C_{T,r}} A_\mu\dd x^\mu\right).
\end{equation}
The coefficient of $T$ in $-\ln\langle\WW\rangle$, after subtraction of the single-source energies, defines the static energy in an auxiliary stationary completion. The pNRQCD matching coefficient is obtained by subtracting the ultrasoft contribution in the same convention \cite{Brambilla2000,Pineda2012}. The auxiliary completion is an infrared regulator for the local static problem.

The relative displacement is a spacelike vector $r^\mu$ in the midpoint tetrad, with $u\cdot r=0$, $r=(r^\mu r_\mu)^{1/2}$, where $u^\mu$ is the tangent to the central geodesic. The invariant
\begin{equation}
 W=C_{\mu\alpha\nu\beta}r^\mu u^\alpha r^\nu u^\beta
   =E_{ij}r^ir^j
\end{equation}
is the trace-free electric Weyl contraction in vacuum. In the parity-even static sector it is the only scalar linear in curvature at the leading anisotropic short-distance order. With the midpoint worldline and separation fixed geometrically, this contraction is invariant under coordinate changes and rotations of the spatial tetrad. A change of the physical observer generally changes the electric Weyl tensor.

Let $S=S_s\mathbf 1_c/\sqrt{N_c}$ and $O=O^a T^a/\sqrt{T_F}$ denote the singlet and octet fields in the conventional color normalization. To leading multipole and static order the terms relevant for the matching are
\begin{align}
 \mathcal L_{\rm pNRQCD}={}&\tr\big[S^\dagger(i\partial_t-V_s)S\big]\nonumber\\
 &+\tr\big[O^\dagger(iD_t-V_o)O\big]\nonumber\\
 &+g V_A\tr\big[O^\dagger r^i\mathcal E_i S
       +S^\dagger r^i\mathcal E_i O\big]+\cdots,
 \label{eftlag}
\end{align}
where $\mathcal E_i$ is the chromoelectric field at the midpoint, $V_A=1$ at leading order, and integration over the relative coordinate and the center coordinate is understood. The curvature dependence of the leading static potential is represented by
\begin{equation}
 \delta\mathcal L=-c_W(r;\mu,\nu)\,\mathcal O_W,
 \qquad
 \mathcal O_W=\frac{W}{r}\tr(S^\dagger S),\label{opW}
\end{equation}
with $\mu$ the coupling-renormalization scale and $\nu$ the soft--ultrasoft factorization scale. The operator is a color singlet. Its leading ultrasoft counterterm is generated by two insertions of the dipole term in Eq.~\eqref{eftlag}. At the accuracy needed for this counterterm, the singlet--octet gap is $\Delta V=V_o-V_s=\CA\as/(2r)$. Curvature corrections to the dipole vertex or gap multiplied by the flat-space self-energy enter at higher perturbative order in the static sector.

\section{Tree-level matching}
\label{tree}
Write the physical Euclidean metric as $g_{\mu\nu}=N^2\tilde g_{\mu\nu}$ with $\dd\tilde s^2=\dd t^2+\tilde\gamma_{ij}\dd x^i\dd x^j$ and $\tilde\gamma_{ij}=N^{-2}\gamma_{ij}$. In four dimensions the classical Yang--Mills and massless Dirac actions and the gauge connection in the loop are Weyl invariant (with the appropriate fermion rescaling). The renormalized Weyl transformation at one loop is treated in Sec.~\ref{weyl}.

The local optical curvature at the midpoint is $\tilde R_{ij}=2E_{ij}$, $\tilde R=0$. For example, the spatial physical Ricci tensor at the midpoint equals $E_{ij}$ by four-dimensional Ricci flatness, and the conformal transformation of the three-metric contributes another $E_{ij}$. The optical geodesic distance between symmetric endpoints is
\begin{align}
 \ell&=\int_{-r/2}^{r/2}\left(1-\tfrac12 E_{ij}n^in^j\lambda^2\right)\dd\lambda
 =r(1-W/24),\label{dist}\\
 \Rz&=\tilde R_{ij}z^iz^j=2W+O(E^2r^4),
\end{align}
where $n^i=r^i/r$ and $z^i$ is the optical separation vector in a parallel frame. The spatial Riemann correction to the length along this radial segment vanishes by antisymmetry. The Van Vleck factor is $U=\Delta^{1/2}=1+\Rz/12$ to this order.

We continue the optical spatial dimension to $d=3-2\eps$, set $b=(d-1)/2=1-\eps$, hold $\Rz$ fixed, and retain $\tilde R=0$ throughout. The bare coupling is $g_B^2=g^2(\mu)(\mu^2e^{\gamma_E}/4\pi)^\eps Z_g^2$. The residual soft pole is subtracted in a tree-normalized dimensionally continued coefficient, with its opposite assigned to the dipole self-energy. In particular the dimensionally continued tree factor and gap must be used consistently before taking $d\to3$. Finite evanescent changes in the continuation or subtraction may shift the Wilson coefficient and ultrasoft reference term oppositely. 

\subsection{Static Coulomb Green function}
The physical static $A_0$ operator is the density-weighted optical scalar Laplacian, $-\partial_i(\sqrt\gamma\,N^{-1}\gamma^{ij}\partial_j)$. With $L=-\tilde\nabla^2$ its Green function away from coincidence is, in three spatial dimensions,
\begin{equation}
 G_1=\frac{U}{4\pi\ell}=\frac1{4\pi r}\left(1+\frac5{24}W\right).
\end{equation}
Multiplication by $-g^2\CF$ gives the leading static interaction,
\begin{equation}
 V_s^{(0)}(r)=-\frac{\CF\as}{r}\left(1+\frac5{24}W\right).
 \label{treepotential}
\end{equation} Smooth boundary-dependent terms are distinct from the local nonanalytic term proportional to $W/r$.

An independent physical-metric check uses the static Maxwell operator $P=-\nabla_i(N^{-1}\nabla^i)$ with the invariant spatial delta function, where the derivatives now use $\gamma$. With $f=\ln N$,
\begin{equation}
 P=N^{-1/2}\big[-\nabla_\gamma^2-\tfrac12\nabla_\gamma^2 f
 +\tfrac14(\nabla_\gamma f)^2\big]N^{-1/2}.
\end{equation}
Since $\nabla_\gamma^2f=0$ at linear order, the local Green function is $N_+^{1/2}N_-^{1/2}$ times the spatial scalar Green function. The physical spatial Ricci tensor is $E_{ij}$ and the radial spatial length is $r$, so
\begin{equation}
 G_{\rm phys}=\frac{(1+W/8)(1+W/12)}{4\pi r}
 =\frac{1+5W/24}{4\pi r}+O(E^2).
\end{equation}
This agrees also with the static spin-independent term of Caicedo and Urrutia \cite{Caicedo2011}: in their Eqs.~(9), (18), (19), (33) and (36), symmetric endpoints give $\mathcal A^{ab}=r^ar^b/12$, $a_{00}=-W/12$, $\xi_h=0$, and $h_{00}^{(1)}=h_{00}^{(2)}=-W/4$. Thus $\tilde a_{00}=5W/12$ and their symmetrized Coulomb factor is $1+5W/24$.

\section{One-loop soft matching}
\label{hard}
The order-$g^4$ static interaction is determined by the vacuum polarization and the connected two-exchange contribution to the logarithm of the Wilson loop. We first compute these diagrams in the optical metric and then restore the physical metric. The coupling is renormalized in the modified minimal subtraction ($\MS$) scheme; the standard flat-space coefficients and coordinate-space logarithm are
\begin{align}
 a_1&=\frac{31}{9}\CA-\frac{20}{9}\TF n_f,\nonumber\\
 \beta_0&=\frac{11}{3}\CA-\frac43\TF n_f,\\
 L_\lambda&=\ln(\lambda r)+\gamma_E,\qquad L_\ell=\ln(\mu\ell)+\gamma_E.\nonumber
\end{align}
Here $\gamma_E$ is Euler's constant.
\subsection{Local propagators and curvature convolution}
\label{prop}
Let $Z=t^2+\ell^2$, $c_b=\Gamma(b)/(4\pi^{b+1})$ and $F=c_bZ^{-b}$. The free Euclidean Feynman-gauge operators are the Hodge Laplacian on one-forms and the scalar Laplacian on ghosts. We employ the local heat-kernel expansion of \cite{Bunch1979,Vassilevich2003,Toms2015,Frob2018}. The optical heat-kernel coefficients give
\begin{align}
 H(t,z)&=UF,\label{H}\\
 D_{ij'}(t,z)&=g_{ij'}H-\frac{c_b}{4(b-1)}\tilde R_{ij'}Z^{1-b},\label{D}\\
 D_{00'}&=H,\quad D_{0i'}=0,\quad D_{\rm gh}=H.
\end{align}
Here $g_{ij'}$ is the parallel propagator. The vector correction follows from the first heat-kernel coefficient $(a_1)_{ij}=\tilde R\tilde\gamma_{ij}/6-\tilde R_{ij}$, whose proper-time integral supplies $\Gamma(b-1)/\Gamma(b)$. These are local parametrices with the dimensionally continued infrared prescription, not full Green functions. Smooth additions are considered in Sec.~\ref{overlap}.

The scalar fractional Green kernels are particularly useful for attaching the Coulomb legs:
\begin{align}
 K_s(x,y)&=\langle x|L^{-s}|y\rangle=k_s\ell^{2s-d}U,\label{frac}\\
 k_s&=\frac{\Gamma(d/2-s)}{4^s\pi^{d/2}\Gamma(s)}.
\end{align}
The formulas hold at linear curvature and neglect derivatives of curvature; powers outside the convergent domain are interpreted as homogeneous distributions. The flat static polarization kernel scales as $\ell^{1-2d}$. Its covariant scalar component corresponds to $L^b$ and becomes $L^{-(2-b)}$ after the two Coulomb legs are attached. Any additional first-order anisotropic kernel can instead be convolved with the flat Coulomb legs.

Since $\Rz$ is a trace-free quadratic polynomial, $\nabla^2\Rz=0$, and
\begin{equation}
 -\nabla^2(\Rz\ell^p)=-p(p+d+2)\Rz\ell^{p-2}.
 \label{harmonic}
\end{equation}
For the isotropic power the last factor is $p+d-2$. Starting from $\Rz\ell^{1-2d}$ and inverting twice with the exponents $1-4b$ and $3-4b$ gives the ratio to the scalar convolution,
\begin{equation}
 \mathcal K(b)=\frac{(-2b)(2-2b)}{(4-2b)(6-2b)}
 =\frac{b(b-1)}{(2-b)(3-b)}.\label{K}
\end{equation}
In particular $\mathcal K(1-\eps)=-\eps/2+O(\eps^2)$. This zero cannot be taken before multiplying the one-loop poles.

\subsection{Light-fermion polarization}
\label{fermion}
The spinor parametrix is obtained by acting with $-\gamma^\mu\tilde\nabla_\mu$ on the spin-parallel scalar Green kernel. The spin connection contributes through the contracted derivative of the spin parallel transporter, $-\gamma^i\tilde R_{ij}z^j/4$; the derivative of $U$ gives $+\gamma^i\tilde R_{ij}z^j/6$. The scalar curvature endomorphism in the squared Dirac operator vanishes here. In a parallel spin frame,
\begin{equation}
 S(t,z)=2bc_b\frac{\gamma^\mu z_\mu}{Z^{b+1}}U
 +\frac{c_b}{12}\frac{\gamma^i\tilde R_{ij}z^j}{Z^b},\label{spin}
\end{equation}
with $z^0=t$ and four-component spinor trace. Contract the static bubble $-g_B^2\TF n_f\tr[\gamma^0 S(x,y)\gamma^0 S(y,x)]$. The reversed separation sign can be factored out of the numerator. The flat numerator is then $4(t^2-\ell^2)(2bc_b)^2/Z^{2b+2}$, and the term linear in the second part of Eq.~\eqref{spin} is $-8(2bc_b)(c_b/12)\Rz/Z^{2b+1}$. Using
\begin{align}
 J_d&=\int_{-\infty}^\infty\frac{\dd t}{Z^d}
 =\frac{\sqrt\pi\Gamma(d-1/2)}{\Gamma(d)}\ell^{1-2d},\label{J}\\
 \int_{-\infty}^\infty\!\frac{t^2-\ell^2}{Z^{d+1}}\dd t&=-\frac{d-1}{d}J_d,
\end{align}
we obtain the integrated bubble curvature factor
\begin{equation}
 1+\Rz\left(\frac16+\frac{d}{24b^2}\right).
\end{equation}
Relative to the scalar-covariantized contribution, which already contains $U$, the extra factor after Coulomb convolution is
\begin{equation}
 \rho_f=\left(\frac1{12}+\frac{d}{24b^2}\right)\mathcal K(b)
 =-\frac5{48}\eps+O(\eps^2).
\end{equation}
With the conventional coordinate-space normalization (per $\TF n_f\as/(4\pi)$ and per dimensionally continued tree exchange) the flat bubble is
\begin{equation}
 f_f=-\frac4{3\eps}-\frac{20}{9}-\frac83 L_\ell+O(\eps),
\end{equation}
Thus the additional finite optical fermion term is
\begin{equation}
 \lim_{\eps\to0}f_f\rho_f\Rz=\frac5{36}\Rz.\label{fextra}
\end{equation}

\subsection{Gluon--ghost polarization}
\label{gluon}
\subsubsection{Static current and Wick contractions}
Split the temporal gauge field as $A_0^a=B^a+a_0^a$, with external $B^a$ static, and set $A_i^a=a_i^a$. In ordinary (not background-field) covariant Feynman gauge, the cubic term linear in $B$ follows from $F_{0i}^2/2$. After integrating the derivative on $B$ by parts it is $g_B\int B^aJ_0^a$, with the measure understood and
\begin{align}
 J_0^a=f^{abc}\big[&-\dot a_i^ba_i^c+2(\tilde\nabla_i a_0^b)a_i^c\nonumber\\
 &+a_0^b\tilde\nabla_i a_i^c\big].\label{current}
\end{align}
The factor of two in the mixed derivative current comes from differentiating the external and the quantum temporal fields. The ghost vertex linear in $B$ has one temporal derivative. The contraction identities $f^{abc}f^{a'bc}=\CA\delta^{aa'}$ and $f^{abc}f^{a'cb}=-\CA\delta^{aa'}$ fix the relative signs. The Hodge identity $\Delta_1\dd=\dd\Delta_0$ gives $\tilde\nabla_x^iD_{ij'}=-\tilde\nabla_{j'}H$ for compatible Green functions, away from contact. The three types of current in Eq.~\eqref{current} are denoted $J_T,J_M,J_D$.

\begin{table}[b]
\caption{Contractions contributing to the static gluon--ghost kernel, after temporal integration by parts. The common color factor $C_A$ and time integral are suppressed.}
\label{wicktable}
\begin{ruledtabular}
\begin{tabular}{lc}
Contraction & Integrand \\\hline
$J_TJ_T$ & $2\dot D_{ij'}\dot D^{ij'}$\\
$J_MJ_M$ & $4D^{ij'}H_{;ij'}$\\
$J_MJ_D$ & $-2|\tilde\nabla_xH|^2$\\
$J_DJ_M$ & $-2|\tilde\nabla_yH|^2$\\
$J_DJ_D$ & $-\dot H^2$\\
Ghost loop & $-\dot H^2$
\end{tabular}
\end{ruledtabular}
\end{table}
The first entry is the sum of a product of first time derivatives and a term with a second time derivative, integrated by parts. The divergence--divergence contraction is $H\partial_t^2 H$ off contact. The ghost loop has the usual Grassmann sign. Combining Table~\ref{wicktable}, the gluon--ghost current kernel divided by its color factor is
\begin{align}
 \mathcal P=\int_{-\infty}^\infty\dd t\,\big[&2\dot D_{ij'}\dot D^{ij'}+4D^{ij'}H_{;ij'}\nonumber\\
 &-2|\tilde\nabla_xH|^2-2|\tilde\nabla_yH|^2-2\dot H^2\big].\label{kernel}
\end{align}
The inverse-propagator insertion is minus the current--current kernel. Ordinary local contact terms are handled with the flat kinetic counterterm and covariant dimensional extension. The trace-free local part of the four-gluon tadpole vanishes since it is proportional to $\tr a_1$; smooth infrared terms have to be treated consistently with the overlap. A cubic vertex with three temporal legs is zero because it cannot arise from the antisymmetric field strength. At this order no other bulk diagrams with only temporal source attachments occur. Spatial attachments are discussed in Sec.~\ref{overlap}.

\subsubsection{Curvature numerator}
Using the bitensor identities collected in App.~\ref{bitensor}, and with primes indicating $Z$ derivatives, one obtains
\begin{align}
 g^{ij'}H_{;ij'}&=U(-2dF'-4\ell^2F'')-\Rz F',\label{derivH}\\
 |\tilde\nabla_xH|^2&=4U^2\ell^2(F')^2+\tfrac23\Rz FF'.\label{derivH2}
\end{align}
The additional spatial-vector correction contracts as $\tilde R_{ij'}F^{;ij'}=-4\Rz F''$. The term in $\dot D_{ij'}\dot D^{ij'}$ linear in the vector endomorphism is trace zero. The curvature-dependent pieces in Eq.~\eqref{kernel}, beyond the common $U^2$ factor, are therefore
\begin{align}
 \delta\mathcal P_{\rm int}&=-4\Rz FF'-\frac83\Rz FF'
 +\frac{4c_b}{b-1}\Rz Z^{1-b}F''\nonumber\\
 &=c_b^2\Rz Z^{-d}\left[\frac{20b}{3}+\frac{4b(b+1)}{b-1}\right].\label{curvnum}
\end{align}
This isolates the derivative geometry from the vector endomorphism; treating each vector component as a scalar would miss the second, dimensionally singular term.

The flat integrated combination reduces to $(2d-6)\dot F^2-4|\nabla F|^2$. Inserting $\int t^2Z^{-d-1}\dd t=J_d/(2d)$ gives the full reduced kernel
\begin{align}
 \mathcal P&=-c_b^2J_d\left[A_b(1+\Rz/6)-B_b\Rz\right],\label{glukernel}\\
 A_b&=4b^2(3+1/d),\qquad B_b=\frac{20b}{3}+\frac{4b(b+1)}{b-1}.
\end{align}
The same $B_b$ follows by performing the derivatives in Eq.~\eqref{kernel} before time integration. The sign convention in Eq.~\eqref{glukernel} is fixed by the momentum-space check in App.~\ref{flatcheck}.

\subsubsection{Flat normalization and anisotropic contribution}
Attaching the Coulomb legs and normalizing to tree exchange gives (per $\CA\as/(4\pi)$):
\begin{align}
 f_{\rm se}&=\frac{\sqrt\pi(3\eps-5)e^{\gamma_E\eps}(\mu\ell)^{2\eps}}{4\Gamma(1/2-\eps)\Gamma(5/2-\eps)}\nonumber\\
 &\qquad\times\Gamma(-\eps)\Gamma(1/2-2\eps)\nonumber\\
 &=\frac5{3\eps}+\frac{31}{9}+\frac{10}{3}L_\ell+O(\eps).\label{fse}
\end{align}
The light-fermion factor obeys $f_f/f_{\rm se}=-8b/(3d+1)$, another check on the flat normalization.

Subtract the curvature already contained in $U$ and use Eq.~\eqref{K} to attach the Coulomb legs. The normalized extra anisotropic factor is
\begin{align}
 \rho_g&=\left(\frac1{12}-\frac{B_b}{A_b}\right)\mathcal K(b)\\
 &=\frac{3\eps^3+24\eps^2-67\eps+36}{12(\eps+1)(\eps+2)(3\eps-5)}\nonumber\\
 &=-\frac3{10}+\frac{497}{600}\eps+O(\eps^2).
\end{align}
The product with Eq.~\eqref{fse} is
\begin{equation}
 f_{\rm se}\rho_g=-\frac1{2\eps}+\frac{25}{72}-L_\ell+O(\eps).\label{glfinite}
\end{equation}
The finite constant $25/72=(5/3)(497/600)-(3/10)(31/9)$ depends on the terms evanescent in the curvature convolution. In particular a direct calculation with $d=3$ before performing the integrals would not determine this number.

An additional local ultraviolet curvature counterterm for two temporal fields at this engineering dimension would be proportional to $\tilde R$ or $\tilde R_{00}$; both vanish in the continuation. Covariant kinetic terms and their ordinary renormalization are retained. 

\subsection{Connected temporal Wilson-line web}
\label{web}
The logarithm of the Wilson loop removes independent repetitions of tree exchange. It does not remove the connected non-Abelian color correlation between two exchanges. For $t>0$ write
\begin{align}
 A(t)&=c_bt^{-2b},\qquad \lambda=1+\Rz/12,\nonumber\\
 B(t)&=c_b\lambda(t^2+\ell^2)^{-b}.
\end{align}
The time propagators for the symmetric and antisymmetric source couplings are $D_\pm=(A\pm B)/2$. Define $Q^a=T_1^a+T_2^a$, $P^a=T_1^a-T_2^a$, with the antiquark generator in the antifundamental. The singlet satisfies $Q^a|s\rangle=0$, and $P^2=4\CF-Q^2$. Acting once with $P^a$ produces an octet. Both crossed and nested four-$P$ color orderings give $4\CF(4\CF-\CA)$; subtracting the disconnected factor $(4\CF)^2$ leaves $-4\CF\CA$. The nested $PQQP$ ordering contributes $+4\CF\CA$.

With $u,v,w\ge0$ the three successive time gaps between ordered vertices, the extensive connected correction is $V_W=-4g_B^4\CF\CA I$ with
\begin{align}
 I=\int_0^\infty\dd u\dd v\dd w\,\big[&D_-(u+v+w)B(v)\nonumber\\
 &-D_-(u+v)D_-(v+w)\big].\label{webI}
\end{align}
The purely same-source part is removed by the single-source subtraction. The mixed $AB$ terms factorize exactly: changing variables in the first term and the two mixed crossed terms gives
\begin{align}
 I_{AB}&=\tfrac12\int_0^\infty\dd s\dd t\,A(s)B(t)
 \big[(s-t)_++\min(s,t)\big]\nonumber\\
 &=\tfrac12\left(\int_0^\infty sA(s)\dd s\right)
 \left(\int_0^\infty B(t)\dd t\right).\label{AB}
\end{align}
The first factor is scaleless in the all-hard dimensional expansion. With a separate infrared regulator its ultraviolet--infrared decomposition must not be discarded; see Sec.~\ref{overlap}.

The remaining opposite-source term integrates to
\begin{align}
 I_{BB}&=-\tfrac12\int_0^\infty\dd v\,B(v)\int_v^\infty\dd t\,t B(t)\nonumber\\
 &=-\frac{c_b^2\lambda^2\ell^{3-4b}}{4(b-1)}
 \frac{\sqrt\pi\Gamma(2b-3/2)}{2\Gamma(2b-1)}.\label{BB}
\end{align}
The nested part contributes a factor $t-v$ under this integral; the crossed part supplies $v$. Normalizing by the tree exchange gives (per $\CA\as/4\pi$)
\begin{align}
 f_W&=-\frac{2\sqrt\pi e^{\gamma_E\eps}(\mu\ell)^{2\eps}\Gamma(-\eps)\Gamma(1/2-2\eps)}{\Gamma(1/2-\eps)^2}\nonumber\\
 &=\frac2\eps+4L_\ell+O(\eps).
\end{align}
Since the web carries $\lambda^2$ and the scalar tree carries $\lambda$, its additional curvature factor is $f_W\Rz/12$. In combination with Eq.~\eqref{glfinite}, the additional optical gluon result is
\begin{equation}
 \CA\Rz\left[-\frac1{3\eps}+\frac{25}{72}-\frac23L_\ell\right].\label{optgl}
\end{equation}
The flat combination $\CA(f_{\rm se}+f_W)+\TF n_f f_f$ reproduces $\beta_0/\eps+a_1+2\beta_0L_\ell$. Subtracting the coupling pole gives the known flat potential. This check constrains the color and normalization conventions, although it does not uniquely determine the curvature calculation.

\subsection{Physical-metric restoration}
\label{weyl}
Classical conformal invariance alone is insufficient to relate the renormalized physical and optical quantum theories. We use the one-loop Weyl Ward insertion \cite{Collins1977} at fixed infrared completion. With $f=\ln N$ and with the coupling absorbed into the gauge connection $\mathcal A=gA$, the gauge-field part of the renormalized action variation is
\begin{equation}
 \delta_f S=-\frac{\beta_0}{32\pi^2}\int\dd^4x\sqrt{\tilde g}\, f\,\mathcal F_{\mu\nu}^a\mathcal F^{a\mu\nu},\label{ward}
\end{equation}
where the sign is fixed by the shift $\ln(\mu\ell)\to\ln(\mu\ell)+f$ for constant $f$. Purely gravitational terms cancel in vacuum-normalized loop expectation values. The loop is renormalized with Weyl-related infrared conditions.

The local normalization of this insertion can be checked before taking the dimensional limit. In $D=4-2\eps$, the physical bare action with the coupling absorbed into the connection is
\begin{equation}
 S_B=\frac{1}{4g_B^2}\int\dd^Dx\sqrt{\tilde g}\,e^{-2\eps f}\mathcal F^2.
\end{equation}
Using $Z_g^2=1-\beta_0g^2/(16\pi^2\eps)+O(g^4)$, the product of $-2\eps f$ and the inverse-coupling counterterm gives $-\beta_0 f\mathcal F^2/(32\pi^2)$, with the same normalization as Eq.~\eqref{ward}. The fermion sector admits the direct check of Sec.~\ref{physferm}.

The static part is governed by $L_f=-\tilde\nabla_i(f\tilde\nabla^i)$, for which
\begin{equation}
 L_f=\tfrac12\{f,L\}+\tfrac12\tilde\nabla^2 f.
\end{equation}
Locally $\tilde\nabla^2 f=0$ at the retained order. Consequently $L^{-1}L_fL^{-1}=\tfrac12(f L^{-1}+L^{-1}f)$ up to terms from the infrared completion. At both symmetric sources $f=W/8$, so Eq.~\eqref{ward} produces $+\beta_0 W/4$ in the normalized one-loop bracket. Equation~\eqref{dist} gives the optical distance correction $-\beta_0 W/12$ from the logarithm. The total restoration is $+\beta_0 W/6$.

Since $\Rz=2W$, the finite extra terms in Eqs.~\eqref{fextra} and \eqref{optgl} become
\begin{align}
 c_A&=2\frac{25}{72}+\frac{11}{18}=\frac{47}{36},\\
 c_f&=2\frac5{36}-\frac29=\frac1{18}.
\end{align}
The remaining optical soft infrared pole, in physical normalization, is
\begin{equation}
 V_{s,{\rm IR}}^{(W)}=+\frac{\CF\CA\as^2W}{6\pi r\eps}.\label{softpole}
\end{equation}
This step identifies the coefficient before ultrasoft subtraction; consistency requires removing this pole with the dipole operator of pNRQCD, not a new ultraviolet counterterm in fundamental QCD.

\section{Ultrasoft renormalization and the matched potential}
\label{running}
The soft pole in Eq.~\eqref{softpole} is the first indication of curvature-dependent ultrasoft running. We now determine the ultraviolet counterterm of the singlet operator directly from the dipole self-energy in pNRQCD.
\subsection{Chromoelectric correlator}
\label{us}
The free chromoelectric correlator in the midpoint frame can be read from Eqs.~\eqref{H} and \eqref{D}. The parallel correlator contains an adjoint temporal Wilson line; the latter is unity at the order needed for the free correlator. With the Euclidean--Lorentzian electric-field sign, define $\mathcal G_{ij}(t)=\partial_t^2D_{ij'}(t,0)-H_{;ij'}(t,0)$ for $t>0$. Explicitly,
\begin{align}
 H_{;ij'}(t,0)&=2bc_b\delta_{ij}t^{-2b-2}-c_b\tilde R_{ij}t^{-2b}/6,\\
 -\partial_t^2D_{ij'}(t,0)&=-2b(2b+1)c_b\delta_{ij}t^{-2b-2}\nonumber\\
 &\quad+\tfrac{2b-1}{2}c_b\tilde R_{ij}t^{-2b}.
\end{align}
Thus its local Laplace spectral reference is
\begin{align}
 \mathcal G_{ij}(t)&=\int_0^\infty\dd\omega\,e^{-\omega t}\mathcal J_{ij}(\omega),\\
 \mathcal J_{ij}(\omega)&=\frac{4b^2c_b}{\Gamma(2b+2)}\delta_{ij}\omega^{2b+1}\nonumber\\
 &\quad-\frac{c_b(b-2/3)}{\Gamma(2b)}\tilde R_{ij}\omega^{2b-1}.
 \label{spectral}
\end{align}
In four spacetime dimensions this becomes $(\delta_{ij}\omega^3-E_{ij}\omega)/(6\pi^2)$. This is a local ultraviolet asymptotic reference, not a globally positive spectral density. In particular the linear curvature term cannot be extrapolated into the low-frequency physical spectrum without additional information.

\subsubsection{Independent physical-metric Hadamard check}
The electric singularity can be checked without the optical endomorphism entering Eq.~\eqref{D}. In the physical Euclidean Ricci-flat metric, the local free vector Green function is $P_{\mu\nu'}/(8\pi^2\sigma)$, where $P$ is parallel transport and $\sigma$ is the world function. The Van Vleck and first logarithmic coefficients vanish at linear curvature after neglecting curvature derivatives. Set $c=1/(4\pi^2)$ and let the endpoints have Euclidean time gap $t>0$ and small spatial offsets $\bm x,\bm y$ from the central worldline. To the required order in the offsets,
\begin{align}
2\sigma&=t^2+|\bm x-\bm y|^2\\
 &\quad+\frac{t^2}{3}E_{kl}(x^kx^l+y^ky^l+x^ky^l)+\cdots,\nonumber\\
P_{00'}&=1+f(\bm x)+f(\bm y)+\cdots,\\
P_{0j'}&=-\tfrac t2 E_{jk}(x+y)^k+\cdots,\\
P_{i0'}&=+\tfrac t2 E_{ik}(x+y)^k+\cdots.
\end{align}
The world-function term is obtained by integrating the physical $g_{00}$ perturbation on the unperturbed segment; the mixed parallel components follow by integrating the physical connection, $\Gamma^0_{0i}=E_{ij}x^j$ and $\Gamma^i_{00}=-E_{ij}x^j$. Spatial-metric corrections to $\sigma$ are quartic in the spatial offsets and do not enter this check. On the central geodesic $P_{ij'}=\delta_{ij}$.

Write $D^{\rm phys}=cP/(2\sigma)$. Taking the four derivatives in the field-strength two-point function, then setting $\bm x=\bm y=0$, gives
\begin{align}
\partial_i\partial_{j'}D^{\rm phys}_{00'}
 &=2c\delta_{ij}/t^4-cE_{ij}/(3t^2),\\
-\partial_i\partial_{0'}D^{\rm phys}_{0j'}
 &=cE_{ij}/(2t^2),\\
-\partial_0\partial_{j'}D^{\rm phys}_{i0'}
 &=cE_{ij}/(2t^2),\\
\partial_0\partial_{0'}D^{\rm phys}_{ij'}
 &=-6c\delta_{ij}/t^4.
\end{align}
The Euclidean field-strength correlator is the sum of these terms. Including the same electric-field Wick-rotation sign as in the optical calculation gives the local positive-frequency reference
\begin{equation}
\mathcal G^{\rm phys}_{ij}(t)=\frac{\delta_{ij}}{\pi^2t^4}
-\frac{E_{ij}}{6\pi^2t^2}+\cdots.
\label{physmaxwell}
\end{equation}
Thus the optical result in Eq.~\eqref{spectral} is reproduced independently in four physical dimensions, including its sign and tensor normalization. This is a check of the residue, not of its dimensionally evanescent terms.

\subsection{Dipole anomalous dimension and leading logarithm}
Using the curvature term in Eq.~\eqref{spectral}, the dipole self-energy is
\begin{equation}
 \Sigma_{\rm US}=-g_B^2\CF r^ir^j\int_0^\infty
 \dd t\,e^{-t\Delta V}\mathcal G_{ij}(t).\label{USdef}
\end{equation}
In $D=4-2\eps$ dimensions the curvature term has time dependence $t^{-2+2\eps}$. Its logarithmic divergence follows from the linear term in the gap expansion, or directly from
\begin{equation}
 \int_0^\infty \frac{\omega^{1-2\eps}\dd\omega}{\omega+\Delta V}
 = (\Delta V)^{1-2\eps}\Gamma(2-2\eps)\Gamma(-1+2\eps).
\end{equation}
Since the gamma-function product begins with $-1/(2\eps)$, the pole is
\begin{equation}
 \Sigma_{{\rm US},{\rm UV}}^{(W)}=-\frac{\CF\CA\as^2 W}{6\pi r\eps}.\label{uspole}
\end{equation}
The counterterm in the singlet potential has the opposite sign. The infrared pole in Eq.~\eqref{softpole} therefore cancels Eq.~\eqref{uspole}. Renormalizing the potential and the dipole self-energy at the same factorization scale yields
\begin{equation}
 \boxed{\nu\frac{\partial c_W}{\partial\nu}
    =\frac{\CF\CA}{3\pi}\as^2+O(\as^3).}\label{RG}
\end{equation}
Equation~\eqref{RG} is the leading ultrasoft mixing of the time-ordered dipole product into the local singlet operator \eqref{opW}. In the same normalization, the flat-space correlator gives a running proportional to $\as r^2(\Delta V)^3$ and hence $O(\as^4/r)$. The Weyl term instead gives a running proportional to $\as W\Delta V$ and hence $O(\as^2 W/r)$. A finite, scale-independent evanescent subtraction at this order cannot change Eq.~\eqref{RG}.

For $r^{-1}\gg\Delta V\gg\max(\sqrt{\|E\|},\Omega_E,\Lambda_{\rm QCD})$ and curvature gradients small on $1/\Delta V$, the local short-time approximation can be used for the leading ultrasoft logarithm in a state with no additional competing scales. The renormalized dipole term supplies $+\CF\CA\as^2W\ln(\Delta V/\nu)/(3\pi r)$; matching the soft coefficient at the inverse separation therefore gives
\begin{equation}
 \boxed{\delta\mathcal E_{W,\log}=
 \frac{\CF\CA\as^2}{3\pi}\frac{W}{r}
 \ln\left(\frac{\CA\as}{2}\right).}\label{energylog}
\end{equation}
This is the curvature analogue of the leading ultrasoft logarithm $O(\as^4\ln\as/r)$ in the flat static energy. The stronger hierarchy is needed for the physical logarithm with the gap as the infrared scale, not for the ultraviolet residue in Eq.~\eqref{RG}.

\subsection{Regulated overlap and spatial connectors}
\label{overlap}
A local parametrix by itself does not fix the smooth infrared part of a propagator. This is especially relevant for the vector endomorphism in Eq.~\eqref{D}, whose proper-time integral is infrared divergent at $b=1$. We isolate the leading sensitivity to such smooth terms rather than assigning a physical meaning to the unregulated propagators at large separation.

\subsubsection{Static-source contour at fixed regulator}
Consider a static completion which agrees with the local geometry in a region of scale $\kappa^{-1}$ and has compatible gauge boundary conditions, no perturbative zero modes relevant to the correlator, and a gap of order $\kappa$. We assume that a perturbative completion can be chosen with $r^{-1}\gg\kappa\gg\max(\Delta V,\sqrt{\|E\|},\Lambda_{\rm QCD})$. These conditions also allow expansions in $\kappa r$ and $\Delta V/\kappa$. The source paths, subtraction and curvature expansion are held fixed in this comparison.

At fixed such regulator, the spatial connectors prepare and annihilate a gauge-invariant static-pair state. If it has nonzero overlap with the lowest singlet state, the spectral representation takes the form $\langle\WW(T,r)\rangle=\sum_n Z_n(r)e^{-T\mathcal E_n(r)}$. Hence
\begin{equation}
 -\ln\langle\WW(T,r)\rangle=T\mathcal E_0-\ln Z_0+O(e^{-\omega_\kappa T}),
\end{equation}
where $\omega_\kappa>0$. Endpoint factors and cusp renormalizations are nonextensive. Equivalently, connected endpoint diagrams cluster into regions of finite time extent at fixed gap, while repetitions of disconnected bulk factors cancel in the logarithm. This argument justifies removing connector overlap effects from the extensive energy under these assumptions, without taking the gap to zero before the long-time limit. 

\subsubsection{Leading smooth contributions}
Decompose the temporal and spatial propagators into their local parametrices and smooth infrared remainders, $H_\kappa=H+\delta H$, $D_\kappa=D+\delta D$. For a symmetric perturbative completion, retain their trace-free curvature-linear pieces at the midpoint:
\begin{equation}
 C_{ij}=\delta D_{ij}(0,0),\qquad M_{ij}=\int_0^\infty\dd t\,t\,\delta H_{;ij'}(t,0).\label{CM}
\end{equation}
One may compare two gapped completions to obtain decaying remainders, or use analytic continuation of these moments for a difference from the local parametrix. Terms at infinity in integrations by parts must vanish in the convergence domain before continuation. In the following calculation the leading mixed hard--infrared, nonanalytic $W/r$ structure is retained; further terms suppressed by $\kappa r$ and analytic all-infrared terms are outside this coefficient.

The only constant trace-free spatial-propagator addition in Eq.~\eqref{kernel} that survives at this order multiplies $4H_{;ij'}$. Inserting the two hard Coulomb legs gives a correction to the Coulomb Green function
\begin{equation}
 \delta G_{\rm se}=4g_B^2\CA C_{ij}\int\frac{\dd^dq}{(2\pi)^d}\frac{q_iq_j e^{i\bm q\cdot\bm r}}{(q^2)^3}.
\end{equation}
Using the exact flat identity
\begin{align}
 C_{ij}\int\frac{\dd^dq}{(2\pi)^d}\frac{q_iq_j e^{i\bm q\cdot\bm r}}{(q^2)^3}
 &=-C_{ij}\partial_i\partial_j(k_3r^{6-d})\nonumber\\
 &=-\tfrac18G_{1,d}(r) C_{ij}r^ir^j,
\end{align}
where $G_{1,d}=k_1 r^{2-d}$, the corresponding singlet term is
\begin{equation}
 \delta V_{\rm se}=g_B^2\CF\Delta V_d\,r^ir^j C_{ij},
 \qquad\Delta V_d=\frac{\CA g_B^2}2G_{1,d}.
 \label{VseIR}
\end{equation}
This identity is dimensionally exact and so also controls the finite part multiplying an evanescent coefficient.

For the time-line web, the same- versus opposite-source propagators for the smooth remainder differ by $\delta A(t)-\delta B(t)=r^ir^j\delta H_{;ij'}(t,0)/2+O(r^4)$. In the leading mixed region of Eq.~\eqref{webI} one exchange is hard and the other has time separation of order $\kappa^{-1}$. Expanding the ordered time factors accordingly gives
\begin{align}
 \delta I_{\rm mixed}&=\frac12\int_0^\infty B_{\rm hard}(v)\dd v
 \int_0^\infty t[\delta A(t)-\delta B(t)]\dd t\nonumber\\
 &=\frac18G_{1,d}r^ir^jM_{ij}.
\end{align}
The separated hard--infrared term of the crossed contribution has additional powers of $\kappa r$; the all-infrared terms do not generate a leading nonanalytic $1/r$ Coulomb factor. Therefore
\begin{equation}
 \delta V_W=-g_B^2\CF\Delta V_d\,r^ir^jM_{ij}.\label{VWIR}
\end{equation}
Combining Eqs.~\eqref{VseIR} and \eqref{VWIR} yields the leading sensitivity $g_B^2\CF\Delta V_d r^ir^j(C_{ij}-M_{ij})$.

In the ultrasoft effective field theory (EFT), the corresponding electric correlator variation is $\delta\mathcal G_{ij}=\partial_t^2\delta D_{ij}-\delta H_{;ij'}$. Taking the first time moment,
\begin{equation}
 \int_0^\infty t\,\delta\mathcal G_{ij}(t)\dd t=C_{ij}-M_{ij},\label{moment}
\end{equation}
by integration by parts with the stated boundary prescription. Expanding Eq.~\eqref{USdef} through first order in $\Delta V_d/\kappa$ leads to the identity
\begin{equation}
 \boxed{\delta V_{\rm se}+\delta V_W=\delta\Sigma_{\rm US}^{(g^4)}.}\label{cancel}
\end{equation}
The cancellation occurs prior to removal of the dimensional regulator and includes the finite terms in the leading smooth overlap.

The explicit proper-time separation check in App.~\ref{propertime} verifies both the pole and the finite cutoff-dependent part of this cancellation.

\subsection{Finite matching and subtraction}
With the pole and the infrared overlap identified, we can specify the finite subtraction and assemble the matched coefficient.
For illustration of the finite convention, extending the local spectral reference in Eq.~\eqref{spectral} over all frequencies and subtracting minimally in the tree-normalized convention at scale $\nu$ gives
\begin{align}
 \Sigma_{\rm ref}^{(W)}&=\frac{g^2\CF W\Delta V}{6\pi^2}
 \left[\ln\frac{\Delta V}{\nu}+\frac12+\ln2\right].\label{usref}
\end{align}
One may derive the constant directly in Euclidean time using $\int_0^\infty t^{-2b}e^{-t\Delta V}\dd t=\Gamma(1-2b)(\Delta V)^{2b-1}$. In this reference expression the gap is held fixed in the tree-normalized bracket; evanescent factors from the tree normalization must be retained consistently when converting the bracket to a dimensionful subtraction. A physical correlator contains a finite remainder relative to Eq.~\eqref{spectral}; Eq.~\eqref{usref} is not a calculation of that remainder. The physical hierarchy for use of this reference was specified above. The auxiliary matching completion in Sec.~\ref{overlap} instead has gap $\kappa\gg\Delta V$ and is used in the matching expansion before regulator removal.

The matched soft logarithm is $-4\CA L_\nu/3$ in Eq.~\eqref{mainA}. The two factorization derivatives cancel:
\begin{equation}
 \nu\partial_\nu V_s^{(W)}=-\nu\partial_\nu\Sigma_{\rm ref}^{(W)}
 =\frac{\CF\CA\as^2 W}{3\pi r}.
\end{equation}
In addition, $\mu\dd\as/\dd\mu=-\beta_0\as^2/(2\pi)+\cdots$ cancels the explicit $L_\mu$ dependence through the displayed order.

Collecting the finite optical polarization and web terms, physical-metric restoration, and the ultrasoft subtraction, we obtain
\begin{align}
 V_s^{(W)}(r;\mu,\nu)&=-\frac{\CF\as(\mu)}{r}W
 \left[\frac5{24}+\frac{\as(\mu)}{4\pi}A_E(\mu,\nu)\right],\label{main}\\
 A_E(\mu,\nu)&=\frac5{24}(a_1+2\beta_0 L_\mu)\nonumber\\
 &\quad+\CA\left(\frac{47}{36}-\frac43 L_\nu\right)
 +\frac{\TF n_f}{18},\label{mainA}
\end{align}
The flat static potential is $-\CF\as[1+\as(a_1+2\beta_0L_\mu)/(4\pi)]/r$. The first line of Eq.~\eqref{mainA} arises from the scalar-covariantized flat correction; the remaining terms require the anisotropic polarization and the connected Wilson-line web. The $\nu$-logarithm agrees with Eq.~\eqref{RG}, and the explicit $\mu$ dependence cancels the running of the tree coupling. The individual contributions to the finite matching coefficient are summarized in Table~\ref{constants}.

\begin{table}[!ht]
\caption{Additional finite terms in the coefficient multiplying $W\alpha_s/(4\pi)$, excluding $5(a_1+2\beta_0 L_\mu)/24$. Color factors are not included in the individual numerical entries. The remaining $-4C_A L_\nu/3$ is separately fixed by ultrasoft subtraction.}
\label{constants}
\begin{ruledtabular}
\begin{tabular}{lcc}
Origin & $C_A$ & $T_Fn_f$\\\hline
Local polarization, $\Rz=2W$ & $25/36$ & $5/18$\\
Connected web (log and pole only) & $0$ & $0$\\
Optical distance and Weyl restoration & $11/18$ & $-2/9$\\
\textbf{Extra finite coefficient} & $\bm{47/36}$ & $\bm{1/18}$
\end{tabular}
\end{ruledtabular}
\end{table}
Equivalently,
\begin{align}
 A_E&=\CA\left(\frac{437}{216}+\frac{55}{36}L_\mu-\frac43 L_\nu\right)\nonumber\\
 &\quad-\TF n_f\left(\frac{11}{27}+\frac59 L_\mu\right).
\end{align}
For $\mathrm{SU}(3)$ and $\mu=\nu=e^{-\gamma_E}/r$ this reduces to $437/72-11n_f/54$. 

To demonstrate cancellation of the factorization scale in a finite expression, combine Eq.~\eqref{main} with the reference self-energy Eq.~\eqref{usref}. In the normalization of $A_E$, their sum has the reference bracket
\begin{align}
 A_{E,\rm ref}^{\rm comb}&=\frac5{12}\beta_0 L_\mu-\frac{11}{27}\TF n_f\nonumber\\
 &\quad+\CA\left[\frac{293}{216}-\frac43\big(\ln(\CA\as)+\gamma_E\big)\right].\label{comb}
\end{align}
The $\ln\as$ term reproduces Eq.~\eqref{energylog}. The displayed constant is the sum with the local spectral reference, rather than with the finite state-dependent remainder of the physical correlator.

\subsection{Finite continuation transformation}
\label{scheme}
Consider a family of trace-free optical continuations with the same four-dimensional endpoint geometry and with $\tilde R_{ij}=2(1+\eta\eps)E_{ij}$ at fixed $W$. The optical distance is held fixed. The four-dimensional tree coefficient and logarithms are unchanged. This family is a useful test because the optical Ricci tensor of a spatial conformal transformation of a $d+1$ dimensional Ricci-flat static geometry would instead be $(d-1)E_{ij}$: the spatial Ricci tensor is $E_{ij}$, and the transformation contributes $(d-2)E_{ij}$. Its optical curvature normalization corresponds to $\eta=-1$. This does not account for the additional nonconformal bare vertices in a fully physical $d+1$ dimensional theory.

All additional polarization and web terms linear in $\Rz$ acquire the factor $(1+\eta\eps)$; the curvature-dependent pole in the normalized soft bracket is $-2\CA/(3\eps)$. In the same tree-normalized scheme the flat ultraviolet pole is canceled before taking the limit. The finite fermionic extra has no remaining pole and is unchanged under this particular deformation. Thus minimal subtraction in the deformed continuation gives
\begin{equation}
 A_E^{(\eta)}=A_E^{(0)}-\frac{2\eta}{3}\CA.
\label{evansoft}
\end{equation}
The normalized reference self-energy has the opposite pole $+2\CA/(3\eps)$ and its electric tensor acquires precisely the same $(1+\eta\eps)$ factor. Define $A_{\rm US}$ by $\Sigma_{\rm ref}^{(W)}=-\CF\as^2W A_{\rm US}/(4\pi r)$. The corresponding finite change is
\begin{equation}
 A_{\rm US}^{(\eta)}=A_{\rm US}^{(0)}+\frac{2\eta}{3}\CA.
\label{evanus}
\end{equation}
Consequently $A_E^{(\eta)}+A_{\rm US}^{(\eta)}=A_{E,\rm ref}^{\rm comb}$ for every $\eta$. For $\eta=-1$, the extra soft gluonic constant is $71/36$ instead of $47/36$, with a compensating $-2\CA/3$ shift in the normalized reference bracket. This illustrates explicitly why the individual soft constant cannot be called continuation-independent. It tests a common curvature normalization change, not arbitrary changes to the local propagator algebra, ultraviolet scheme, source operators or evanescent vertices.

The finite non-Abelian constant is quoted in ordinary Feynman gauge with the optical continuation defined in Sec.~\ref{tree}; the closed Wilson loop and the adjoint-linked electric correlator are gauge invariant in a compatible regulator. The continuation test concerns the common trace-free curvature rescaling, while the physical-metric light-flavor calculation below tests a different set of vertices.

\section{Physical-metric check of the light-flavor term}
\label{physferm}
A separate check of the fermion bubble follows from the physical Dirac parametrix. We continue the physical vacuum condition $R_{\mu\nu}[g]=0$ in $D=d+1$ dimensions. The corresponding optical relation is $\tilde R_{ij}=(d-1)E_{ij}$ rather than $2E_{ij}$ at nonzero $\eps$. The following calculation retains the physical evanescent Maxwell kinetic operator.

The Lichnerowicz formula, Ricci flatness, and the algebraic Bianchi identity imply, to linear curvature without curvature derivatives, that the contracted derivative of the spin parallel propagator has no curvature contribution: $\gamma^\mu R_{\mu\nu ab}\gamma^{ab}$ reduces to a Ricci contraction. The spinor Green function thus has the form
\begin{equation}
 S_g(x,y)=2bc_b\frac{\gamma_\mu v^\mu}{(s_g^2)^{b+1}}P_{\rm spin}(x,y),
 \qquad v^\mu=\nabla_x^\mu\sigma_g,
\end{equation}
where $s_g^2=2\sigma_g$. The Van Vleck factor is unity at this order in physical Ricci-flat geometry. For the unit static time directions $u_x,u_y$ and their parallel transport, the spinor trace (after removing the reversed separation sign) is
\begin{align}
 \tfrac14\tr(\slashed u_x\slashed v\slashed u_y^{P}\slashed v)
 &=2(u_x\cdot v)(u_y^P\cdot v)-(u_x\cdot u_y^P)v^2\nonumber\\
 &=2p_t^2/(N_xN_y)-s_g^2,
\end{align}
where $p_t$ is the conserved geodesic momentum associated with the frozen time translation and $u_x\cdot u_y^P=1+O(E^2)$.

Write $f=\ln N$ and let $\bar f$ be its average along the zeroth-order optical segment, with $Z=t^2+\ell^2$. Varying the geodesic action gives $s_g^2=Z(1+2\bar f)$ and $p_t=t(1+2\bar f)$. For the quadratic lapse,
\begin{equation}
 Q_f=f_x+f_y-2\bar f=\frac{E_{ij}z^iz^j}{6}.
\end{equation}
The two color-current vertices and volume factors contribute $N_x^dN_y^d$ relative to the optical measures. Expanding the trace times the Green-function denominators gives the integrand (up to its flat normalization)
\begin{equation}
 \frac{2t^2-Z+Q_f[d(2t^2-Z)-2t^2]}{Z^{d+1}}.
\end{equation}
Using the integrals in Sec.~\ref{fermion}, the relative static polarization factor becomes
\begin{equation}
 1+\frac16\left(d+\frac1{d-1}\right)E_{ij}z^iz^j.\label{physfkernel}
\end{equation}
The curvature factor from the optical spinor kernel equals Eq.~\eqref{physfkernel} after substituting $\tilde R_{ij}=(d-1)E_{ij}$. This checks the curvature numerator at continued $d$, not only at $d=3$.

The extra anisotropic factor relative to the scalar Van Vleck term, in units of $E_{ij}z^iz^j$, is
\begin{align}
 P_f(d)&=\frac16\left(d+\frac1{d-1}\right)-\frac{d-1}{12}\nonumber\\
 &=\frac5{12}-\frac{\eps}{12}+O(\eps^2).
\end{align}
Multiplying by $\mathcal K(b)=-\eps/2+O(\eps^2)$ and the flat fermionic pole $-4/(3\eps)$ gives the additional finite term $5\TF n_f/18$ in units of $W$.

The physical dimensionally continued Maxwell action contains the static density $\sqrt{g}g^{00}g^{ij}=N^{-2\eps}\sqrt{\tilde\gamma}\tilde\gamma^{ij}$. Its operator is $P=L-2\eps L_f$, $L_f=-\tilde\nabla_i(f\tilde\nabla^i)$, to the required order. With $G=L^{-1}$ and $\tilde\nabla^2 f=0$ locally,
\begin{equation}
 G_P=G+\eps(fG+Gf)+O(\eps^2,E^2).
\end{equation}
Because the divergent part of the massless Dirac bubble is proportional to $L$, the two external legs supply $G_PLG_P$, whereas the coupling counterterm supplies $G_P$. Their difference times $\beta_{0,f}/\eps$ leaves
\begin{equation}
 \frac{\beta_{0,f}}{\eps}(G_PLG_P-G_P)=\beta_{0,f}(fG+Gf),
\end{equation}
where $\beta_{0,f}=-4\TF n_f/3$. At the symmetric endpoints $f_++f_-=W/4$, so the leg correction is $-\TF n_f/3$ in units of $W$. The distance conversion supplies $-\beta_{0,f}/12=\TF n_f/9$. Thus the net additional term is
\begin{equation}
 \TF n_f\left(\frac5{18}-\frac13+\frac19\right)=\frac{\TF n_f}{18}.
\end{equation}
Including the scalar-covariantized flat-space term, the full light-flavor bracket is $-\TF n_f(11/27+5L_\mu/9)$, in agreement with Eq.~\eqref{mainA}. This calculation uses the physical Ricci-flat spinor propagator and evanescent Coulomb legs, with the same local matching and coupling-renormalization convention.
\section{Discussion}
\label{discussion}
The flat-space and curvature-dependent ultrasoft logarithms have different dependences on the singlet--octet gap. Their relative magnitude is governed parametrically by $W/\alpha_s^2$. In the perturbative adiabatic regime, however, the curvature scale is small compared with the gap, and the curvature logarithm remains a perturbative correction.

The ultraviolet residue is independent of the smooth part of the chromoelectric correlator. The finite ultrasoft contribution generally depends on the state and the background geometry.

For Coulombic heavy quarkonium, the curvature correction to the static potential must be considered together with the curvature dependence of the kinetic operators. The static coefficient obtained here enters the matching of the corresponding finite-mass theory.

\section{Conclusion}
We have obtained the static singlet potential through one loop and to linear order in the electric Weyl tensor in the specified subtraction scheme. The curvature dependence of the chromoelectric correlator gives rise to ultrasoft renormalization at order $\alpha_s^2$ in the curvature sector. The corresponding anomalous dimension determines the leading curvature logarithm in the static energy. The soft infrared pole and the ultrasoft ultraviolet pole cancel, and the finite light-flavor contribution is reproduced independently in the physical metric.

\section*{Data Availability}

No data were created or analyzed in this study.

\appendix
\section{Bitensor identities and flat-space normalization}
\subsection{Local derivative identities}
\label{bitensor}
The bitensor derivatives used in Eq.~\eqref{kernel} are, in the common parallel frame,
\begin{align}
 g^{ij'}\sigma_{;ij'}&=-d-\Rz/6,\quad\sigma=\ell^2/2,\\
 \tilde\nabla_xU&=\tilde R\cdot z/6,\quad \tilde\nabla_yU=-\tilde R\cdot z/6,\\
 g^{ij'}U_{;ij'}&=-\tilde R/6=0.
\end{align}

\subsection{Flat normalization from momentum space}
\label{flatcheck}
An independent check is available from the flat ordinary-gauge three-gluon vertex and Faddeev--Popov loop. At static momentum $q=(0,\bm q)$, the contracted numerator including the gluon symmetry factor and ghost subtraction is
\begin{equation}
 (2d-2)k_0^2+\tfrac12|\bm k-\bm q|^2+\tfrac12|\bm k+2\bm q|^2.
\end{equation}
The gluon-only numerator has $(2d-1)k_0^2$; the ghost loop subtracts $k_0^2$. In $d+1$ spacetime dimensions the bubble integral is
\begin{align}
 I_0(q)&=\int\frac{\dd^{d+1}k}{(2\pi)^{d+1}}\frac1{k^2(k+q)^2}\\
 &=\frac{(q^2)^{-\eps}}{(4\pi)^{2-\eps}}\frac{\Gamma(\eps)\Gamma(1-\eps)^2}{\Gamma(2-2\eps)}.
\end{align}
Using $\int k\cdot q/[k^2(k+q)^2]=-q^2I_0/2$ and $\int k_0^2/[k^2(k+q)^2]=-q^2I_0/(4d)$, with the same integration measure, the contracted bubble is $(3d+1)q^2I_0/(2d)$. Fourier transformation gives Eq.~\eqref{fse}.

\section{Auxiliary regulator integrals and dimensional expansions}
\subsection{Proper-time separation}
\label{propertime}
One can explicitly test the moment identity by splitting the local heat-kernel integral at $s=S=\kappa^{-2}$. This is an auxiliary separation of the parametrix, not by itself a gauge-invariant regulator of the complete theory. Removing the tail $s>S$ from the vector endomorphism changes the coincident propagator by
\begin{align}
 C_{ij}&=\frac{\tilde R_{ij}}{(4\pi)^{b+1}}\int_S^\infty s^{-b}\dd s
 =\frac{\tilde R_{ij}}{(4\pi)^{b+1}}\frac{S^{1-b}}{b-1},\\
 M_{ij}&=\frac{\tilde R_{ij}}{3(4\pi)^{b+1}}\frac{S^{1-b}}{b-1}.
\end{align}
Here $M_{ij}$ follows from the mixed derivative of $U$ and $\int_0^\infty t e^{-t^2/(4s)}\dd t=2s$. The tails are evaluated for $\mathrm{Re}\,b>1$ and continued afterwards. Thus
\begin{equation}
 C_{ij}-M_{ij}=-\frac{4 E_{ij}}{3(4\pi)^{2-\eps}}\frac{\kappa^{-2\eps}}{\eps}.
\end{equation}
Normalized to the dimensionally continued tree exchange, the additional one-loop bracket is
\begin{align}
 A_{\rm ov}&=\frac{2\CA}{3\eps}e^{\gamma_E\eps}\left(\frac{\mu}{\kappa}\right)^{2\eps}\nonumber\\
 &=\CA\left[\frac2{3\eps}+\frac43\ln\frac{\mu}{\kappa}+\frac23\gamma_E\right]+O(\eps).
\end{align}
This explicitly reproduces the opposite of the soft infrared pole in the bracket, together with its finite separation constant. Subtracting the EFT overlap renormalized at $\nu$ removes both $\ln\kappa$ and the auxiliary $2\gamma_E/3$ constant. This gives Eq.~\eqref{mainA} in the prescription used here.

\subsection{Laurent-series identities}
We collect the Laurent-series inputs for the finite coefficient (each loop normalized by the dimensionally continued tree Green function and its own color factor):
\begin{align}
 f_{\rm se}&=5/(3\eps)+31/9+(10/3)L_\ell,\\
 f_W&=2/\eps+4L_\ell,\\
 f_f&=-4/(3\eps)-20/9-(8/3)L_\ell,\\
 \rho_g&=-3/10+(497/600)\eps,\\
 \rho_f&=-(5/48)\eps.
\end{align}
Only the displayed orders are required in the products. The gluon residue is $-1/2+2/12=-1/3$ per $\Rz$ and the finite part is $25/72$. The fermion residue vanishes because $\rho_f=O(\eps)$ while the product has finite part $5/36$. The isotropic pole is $5C_A/3+2C_A-4T_Fn_f/3=\beta_0$. The ordinary coupling renormalization removes that scalar-covariantized pole, including the tree factor $U$. The residual infrared pole is treated separately, using the same dimensionally continued tree normalization in the soft and ultrasoft sectors.

The short-distance curvature calculation is performed at $O(E)$, retaining the nonanalytic $Er$ behavior. Curvature derivatives, higher multipoles and powers of curvature are not included. The numerical constants quoted here are exact rational numbers within this expansion.

\end{document}